\documentclass[12pt]{article}
\usepackage{amsfonts,epsfig,latexsym,amscd}
\usepackage{amssymb}

\newcommand{\I}{{\mathbb{I}}}

\newcommand{\be}{\begin{equation}}
\newcommand{\ee}{\end{equation}}
\newcommand{\bea}{\begin{eqnarray}}
\newcommand{\eea}{\end{eqnarray}}
\newcommand{\bean}{\begin{eqnarray*}}

\newcommand{\eean}{\end{eqnarray*}}
\font\upright=cmu10 scaled\magstep1

\newcommand{\PP}{\hbox{\upright\rlap{I}\kern 1.5pt P}}

\newcommand{\identity}{{\upright\rlap{1}\kern 2.0pt 1}}

\newcommand{\HH}{\mbox{\hbox{\upright\rlap{I}\kern 1.7pt H}}}

\newcommand{\fr}{\frac}

\newcommand{\ra}{\rightarrow}
\newcommand{\al}{\alpha}
\newcommand{\sg}{\sigma}
\newcommand{\bt}{\beta}
\newcommand{\pr}{\partial}
\newcommand{\x}{ {\bf x} }
\newcommand{\hs}{\hspace{5mm}}

\newcommand{\dg}{\dagger}

\newcommand{\ve}{\varepsilon}
\newcommand{\acc}{\\[3mm]}

\newcommand{\vv}{{\bf v}}

\begin{document}
\setcounter{page}{0}
\begin{titlepage}
\strut\hfill
\vspace{0mm}
\begin{center}

{\large\bf Solutions of the Generic Non-Compact Weyl Equation}
\vspace{12mm}

{\bf Anastasia Doikou${}^*$ \ and \ Theodora Ioannidou${}^\dg$}
\\[8mm]
\noindent ${}^*${\footnotesize Department of Engineering Sciences, University of Patras,
GR-26500 Patras, Greece }\\
{\footnotesize {\tt E-mail: adoikou@upatras.gr}}
\\[8mm]
\noindent ${}^\dg${\footnotesize Department of Mathematics, Physics and Computational Sciences, Faculty of Engineering,\\
Aristotle University of Thessaloniki, GR-54124 Thessaloniki, Greece }\\
{\footnotesize  {\tt E-mail: ti3@auth.gr}}

\vspace{12mm}

\begin{abstract}
\noindent
In this paper, solutions of the generic  {\it non-compact} Weyl equation are obtained.
In particular, by identifying a suitable similarity transformation and introducing a non-trivial change of variables we are able to  implement azimuthal dependence on the solutions of the {\it diagonal non-compact} Weyl equation derived in \cite{DI3}.
We also discuss some open questions related to the construction of infinite BPS monopole configurations.

\end{abstract}

\end{center}
\end{titlepage}

\section{Introduction}
Direct construction of BPS monopole configurations  with monopole number greater than one is a  very difficult task.
One way to bypass this difficulty is the  so-called {\it inverse Nahm transform}.
In this approach a nonlinear ordinary differential equation (i.e. the Nahm equation) must be solved and its solutions are used to define the Weyl equation \cite{Nahm}.
Then, the monopole fields can be constructed from the solutions of the Weyl equation.

The  Nahm equations provide a system of non-linear ordinary differential equations of the form
\be \fr{dT_i}{ds}=\fr{1}{2}\, \ve_{ijk}\,[T_j,\ T_k],
\label{Nahm}
\ee
where $T_i$ are complex-valued functions of the variable $s$, known as  {\it Nahm data}; and $\ve_{ijk}$ is the totally antisymmetric tensor.
For $SU(n+1)$ spherically symmetric BPS monopoles with  minimal symmetry breaking case,   the Nahm data $T_i$'s can be cast as
(for a  detailed discussion, see Ref. \cite{msbook})
\be
T_i (s)=-{i\over 2}\,f_i(s)\,\tau_i, \hs i=1,\ 2,\ 3 \label{data}
\ee
where $\tau_i$'s form  the $n$-dimensional representation of $SU(2)$ and satisfy:
\be
[\tau_i,\ \tau_j] = 2i \varepsilon_{ijk}\,\tau_k.
\ee
Given the Nahm data for a $n$-monopole the one-dimensional Weyl equation
\be
\left(\fr{d}{ds}-\I\otimes x_j \sg_j +iT_j\otimes\sg_j\right)\vv({\bf x},s)=0
 \label{Weyl}
\ee
for the complex $2n$-vector $\vv(\x,s)$, must be solved.
${\bf x}=(x,y,z)$ is the position in space at which the monopole fields are to be calculated and $\sigma_i$'s are the familiar Pauli matrices.

Recently, a non-compact approach of the inverse Nahm transform was proposed in \cite{DI3} by introducing an infinite dimensional spin representation of the $\mathfrak{sl}_2$  algebra for the Nahm data.
In this case, the  $\sigma_i$'s become the two dimensional spin half representation of $\mathfrak{su}(2)$, while the Nahm data were given in terms of appropriate differential operators.
Thus, the Weyl equation was written in terms of differential operators, to include infinite dimensional representations of $\mathfrak{sl}_2$, and not in terms of $n\times n$ matrices as in its conventional form \cite{DI,DI2}.
The corresponding equation is called the {\it non-compact} Weyl equation.
Finally, the equivalence between the matrix versus the differential operator description of the Weyl equation via the spin representation of  $\mathfrak{su}_2$ was studied in detailed.

In this formalism,  $f_i(s)=\fr{1}{s}$ due to the minimal symmetry breaking  while the representations $\tau_i$'s are given by the spin $S$ representation of $\mathfrak{sl}_2$ algebra:
\be
\tau_1 =\! -\!\left(\xi^2 -1\right){d\over d\xi} + S \left(\xi +\xi^{-1}\right), \ \ \tau_2 = - i\left [\left(1+\xi^2\right) {d\over d\xi} + S\left(\xi^{-1} -\xi\right) \right ], \ \  \tau_3= -2\xi\, {d\over d\xi}. \label{rep1}
\ee
Note that, for $S$ being an integer or half integer one deals with the $n$-dimensional\footnote{Where $n=2S+1$ and $S\neq 0$.} representation of ${\mathfrak su}(2)$.
This leads to the so-called {\it differential finite} Weyl equation, which is naturally equivalent to the conventional Weyl equation as shown  in the Appendix of \cite{DI3}.
In  the basis of polynomials of $\xi$ on the unit circle (i.e. $\xi = e^{i\al}$), the inner product takes  the form:
\be
\langle f, g \rangle= {1\over 2i\pi} \int {1\over \xi} \,f^* g\, d\xi \label{inner}.
\ee
Then $\langle \xi^m, \xi^n \rangle = \delta_{nm}$.

Similarly to $\tau_i$'s given by (\ref{rep1}), the $\sigma_i$'s are expressed in terms of the spin half  representation of the  variable $\eta$  as
\be
\sigma_1=\! -\!\left(\eta^2-1\right) {d \over d\eta} + \fr{ \left(\eta^{-1} +\eta\right)}{2}, \ \
\sigma_2 = \!-i\left [\left(1+\eta^2\right) {d \over d\eta} + \fr{\left(\eta^{-1}-\eta\right)}{2}\right], \ \ \sigma_3 =-2\eta\, {d\over d\eta}. \label{ss}
\ee

Then the  Weyl equation (\ref{Weyl}) takes its non-compact form:
\bea
&&\!\!\!\!\!\!\!\!\!\!\!\!\!\!\!\!\!\!\!\! \left\{ {d\over d s} - {1\over 2s} \left [(\xi^2-1) {d \over d \xi} -S \left(\xi +\xi^{-1}\right)\right ] \!\!\!\left [ \left(\eta^2 -1\right){d\over d \eta} -\fr{\left(\eta^{-1} + \eta\right)}{2}\right ] \right. \nonumber\acc
&&\!\!\!\!\!\!\!\!\!\!\!\!\!\!\!\!\!\!\!\! \left. +{1\over 2s} \left [\left(\xi^2 +1\right) {d \over d \xi} +S \left(\xi^{-1} - \xi\right)\right ]\!\!\!\left [\left(\eta^2 +1\right) \fr{d}{d\eta}+\fr{\left(\eta^{-1} -\eta\right)}{2} \right ] \!\! -{2 \xi \eta\over s} {d^2 \over d \xi\,d\eta}\right.\nonumber\acc
&&\!\!\!\!\!\!\!\!\!\!\!\!\!\!\!\!\!\!\!\!\left.+ x\left[(1-\eta^2)\fr{d}{d\eta}+\fr{\eta+\eta^{-1}}{2}\right]-iy\left[(\eta^2+1)\fr{d}{d\eta}+\fr{\eta^{-1}-\eta}{2}\right]-2z \eta {d \over d \eta} \right \} \vv\left({\bf x},\eta,\xi,s\right)=0,  \label{ww}
\eea
provided that $S=0$ or {\it not} an integer or half integer.

Let us first concentrate on the differential finite $SU(n+1)$ case. In order to construct the monopole fields, first  choose  together with the inner product (\ref{inner}) an appropriate orthonormal basis
$\{\hat{\vv}_1,\dots,\hat{\vv}_{n+1}\}$   for $n=2S+1$ of the differential finite Weyl equation (\ref{ww}), satisfying
\be
\int_0^{n+1}\,\langle \hat{\vv}_i,\hat{\vv}_j \rangle  \,ds = \delta_{ij}.\label{cod}
\ee
Then  the Higgs field $\Phi$ and the gauge potentials $A_k$ for $k=1,2,3$ are given by
\bea
\Phi_{ij}&=& -i \int_0^{n+1} (s-n) \,\langle \hat{\vv}_i,\hat{\vv}_j \rangle  \,ds, \nonumber\acc
A_k&=&\int_0^{n+1}  \langle \hat{\vv}_i,\pr_k\hat{\vv}_j \rangle \, ds. \label{fields}
\eea

In our earlier work \cite{DI3}, solutions of the non-compact Weyl equation  (\ref{ww}) were obtained explicitly for the infinite dimensional spin zero representation of $\mathfrak{sl}_2$ in the {\it diagonal case}; i.e. when ${\bf x}=(0,0,r)$.
In that case, the range of the parameter $s$  in order to avoid divergencies of the solutions is $s\in(-\infty,1]$ while   in addition to the inner product (\ref{inner}) an appropriate {\it infinite} dimensional orthonormal basis $\{\hat{\vv}_1,\dots,\hat{\vv}_{n}\}$, i.e. for $n\ra \infty$ was in demand.
The associated solutions were expressed in terms of the Kummer functions.
Then a suitable infinite set of orthogonal functions were chosen, and in analogy to the finite case, expressions of the relevant Higgs fields were proposed. These expressions turned out to have a simple and elegant form, and should correspond to a kind of infinite spherically symmetric BPS monopole configurations. However, the explicit form of the corresponding monopole configurations and their properties like their energy and topological charge need further careful investigation.
Although, it was explicitly shown that in the finite case the two formulations --matrix versus finite differential form-- of the Weyl equation are equivalent, in the non-compact case the picture changes drastically. It might be the case that even the corresponding Bogomolny equations are not the usual ones at this limit and they could be related to the $SU(\infty)$  Yang-Mills equations, and in general to the large-$N$ limit of $SU(N)$ theories \cite{R}. Then we will be able to have a better understanding of the corresponding configurations and compare them with the finite case.

The next natural step is to verify that our results satisfy the Bogomolny equation. However, in order to do so we need to implement azimuthal dependence to the radial solutions obtained in \cite{DI3}.
That way, solutions of the generic non-compact  Weyl equation given by (\ref{ww}) can be derived.
A similar approach has been applied in \cite{DI2} in order to construct $SU(n+1)$ spherically symmetric monopole solutions of the conventional full Weyl equation in the case of the minimal symmetry breaking. Following the same methodology in the non-compact case, we seek for a suitable transformation \cite{DI2} that reduces the full problem to the {\it diagonal} one. That way generic solutions of the non-compact Weyl equation  can be constructed from the known radial ones.
This approach is described  in the next section.

\section{Azimuthal Dependence}

For consistency reasons we keep the same terminology used in our paper  \cite{DI2}.
There the conventional Weyl equation was identified as a Hamiltonian system containing some bulk spin-spin interaction and a boundary term.   For that reason,  the  second term of  (\ref{Weyl}) was called  {\it boundary term} and the third term of (\ref{Weyl}) was called  {\it bulk term}.

In analogy,
the aforementioned terms in the non-compact case (\ref{ww}) take the form
\bea
\I\otimes x_i\sigma_i\!\!&=&f(\eta)\fr{d}{d\eta}+g(\eta)\label{bd},\acc
iT_i\otimes\sigma_i&=&\fr{(\eta-\xi)^2}{s}\fr{d^2}{d\eta d\xi}-\fr{1}{2s}\fr{\eta^2-\xi^2}{\eta}\fr{d}{d\xi}+\fr{S}{s}\fr{\eta^2-\xi^2}{\xi}\fr{d}{d\eta}-
\fr{S}{2s}\fr{\eta^2+\xi^2}{\eta\xi},\label{bu}
\eea
where  the functions $f(\eta)$ and $g(\eta)$   are equal to
\bea
f(\eta)&=&-w\left(\eta^2+\fr{2z}{w}\eta-\fr{\bar{w}}{w}\right), \label{f}\acc
g(\eta)&=&\fr{1}{2}w\eta^{-1}\left(\eta^2+\fr{\bar{w}}{w}\right), \label{g}
\eea
and  $w=x+iy$. Using spherical coordinates the variables become: $w=r\sin\theta\, e^{i\phi}$ and $z=r\cos\theta$.
In \cite{H}, Hitchin obtained a direct correspondence between monopoles and bundles over the mini-twistor space ${\mathrm \Pi}$, which is a $2$-dimensional complex manifold isomorphic to the holomorphic tangent bundle to the Riemann sphere T${\mathbb C}$P$^1$.
The same coordinates appear in our formulation.
In particular, the coordinates on $\Pi$  is the standard inhomogeneous coordinate on the base space defined by $\eta$  and   the complex fibre coordinate defined by  $f(\eta)$. For the twistor transform these twistor coordinates are related to the space coordinates $x,y,z$ via equation (\ref{f}).

Next, as in \cite{DI2}, we need to derive a transformation which {\it diagonalizes} the boundary term  (\ref{bd}).
Since we are dealing with differential operators this can be achieved by introducing a function ${\cal F}(\eta)$ such that:
\be
\fr{1}{{\cal F}(\eta)}\left(\I\otimes x_i\sigma_i\right){\cal F}(\eta)=f(\eta)\fr{d}{d\eta}.\label{ff}
\ee
In accordance a new variable $\pi$ needs to be introduced  in order to  diagonalize what is left of the boundary  term, ie
\be
f(\eta)\fr{d}{d\eta}\ra -2r\pi\fr{d}{d\pi}.\label{cc}
\ee
From (\ref{ff}) one can easily obtain that
\be
{\cal F}(\eta)=\sqrt{\fr{\left(\eta+\fr{z+r}{w}\right)\left(\eta+\fr{z-r}{w}\right)}{\eta}}.
\label{F}
\ee
Similarly, (\ref{cc}) implies the following change of variables
\be
\eta=\fr{\pi\left(r-z\right)+r+z}{w \left(\pi-1\right)}.\label{eta}
\ee

Finally, the bulk term has to be invariant under the aforementioned transformation.
This implies the existence of another function ${\cal G}(\xi)$, which on the one hand leaves
the boundary term invariant and on the other hand satisfies the following condition:
\be
\fr{1}{{\cal G}(\xi){\cal F}(\eta)}\left(iT_i\otimes\sigma_i\right){\cal F}(\eta){\cal G}(\xi)=iT_i\otimes\sigma_i.\label{bb}\ee
It is straightforward to show that, in this case, the function $G(\xi)$ is given by
\be
{\cal G}(\xi)=\left[\fr{\left(\xi+\fr{z+r}{w}\right)\left(\xi+\fr{z-r}{w}\right)}{\xi}\right]^S.
\label{G}
\ee
Note that the form of the function ${\cal G}(\xi)$ is a generalized version of the function  ${\cal F}(\eta)$ and holds for any $S$. More precisely, ${\cal F}(\eta)\equiv{\cal G}(\xi){\Big |}_{S=\fr{1}{2}}^{\xi\ra \eta}$.

In order to show that the condition (\ref{bb}) is satisfied  observe that  the bulk term transforms as
\bea
\!\!\fr{1}{{\cal G}(\xi){\cal F}(\eta)}\left(iT_i\otimes\sigma_i\right){\cal F}(\eta){\cal G}(\xi)&\!\!=\!\!&\fr{\left(\eta-\xi \right)^2}{s}\fr{d^2}{d\eta d\xi}+\fr{1}{s}\left(\left(\eta-\xi\right)^2\fr{d\ln{\cal F}}{d\eta}-\fr{\eta^2-\xi^2}{2\eta}\right)\fr{d}{d\xi}\nonumber\acc
&&\!\!+\fr{1}{s}\left(\left(\eta-\xi\right)^2\fr{d\ln{\cal G}}{d\xi}+\fr{S\left(\eta^2-\xi^2\right)}{\xi}\right)\fr{d}{d\eta}
\nonumber\acc
&&\!\!+\fr{\left(\eta-\xi \right)^2}{s}\fr{d\ln{\cal F}}{d\eta}\fr{d\ln{\cal G}}{d\xi}-\fr{S}{2s}\fr{\eta^2+\xi^2}{\eta\xi}\nonumber\acc
&&\!\!-\fr{1}{2s}\fr{\eta^2-\xi^2}{\eta}
 \fr{d\ln{\cal G}}{d\xi}+\fr{S}{s}\fr{\eta^2-\xi^2}{\xi}\fr{d\ln{\cal F}}{d\eta}.
 \label{tr}
\eea
However, by introducing the new  variables $\pi$ and $\rho$ instead of $\eta$ and $\xi$, respectively, given by (\ref{eta})
\be
\pi=\fr{\eta+\fr{z+r}{w}}{\eta+\fr{z-r}{w}},\hs\hs \hs \rho=\fr{\xi+\fr{z+r}{w}}{\xi+\fr{z-r}{w}},
\label{ph}
\ee
it  can be easily  verified that (\ref{tr}) gives (\ref{bb}).

{\bf Remark:} The condition (\ref{bb}) is satisfied even at the spin zero representation, that is, for $S=0$. Then  ${\cal G}(\xi)=1$ and the only necessary requirement is the change of variables.

When these transformations act on the generic non-compact Weyl equation (\ref{ww})  the last takes the simple form
{\small  \be
\left( {d\over d s} -\fr{(\pi-\rho)^2}{s}\fr{d^2}{d\pi d\rho}+\fr{1}{2s}\fr{\pi^2-\rho^2}{\pi}\fr{d}{d\rho}-\fr{S}{s}\fr{\pi^2-\rho^2}{\rho}\fr{d}{d\pi}+\fr{S}{2s}
\fr{\pi^2+\rho^2}{\pi\rho}-2r\pi\fr{d}{d\pi}\right)\vv_0({\bf x},\pi,\rho,s)=0 \label{ssW}\ee}
which is nothing else but the diagonal  (ie. when ${\bf x}=(0,0,r$))  non-compact Weyl equation  introduced in \cite{DI3}.
Equation (\ref{ssW}) has been studied in detail and its solutions have been obtained in terms of hypergeometric functions.

Let $\vv_0({\bf x}, \pi,\rho,s)$ be the solution of the diagonal case (\ref{ssW}). Then the solution $\vv\left({\bf x},\eta,\xi,s\right)$ of the generic problem (\ref{ww}) can be obtained from the radial solution  $\vv_0$   via the relation
\be
\vv\left({\bf x},\eta,\xi,s\right)={\cal F}(\eta)\ {\cal G}(\xi)\ \vv_0({\bf x}, \pi,\rho,s).\label{gen}
\ee
Note that we also need to impose the new variables defined in (\ref{ph}). In the Appendix we present the corresponding results for the   $SU(n+1)$ finite case, and discuss their relevance with the results obtained in the conventional case \cite{DI2}.

We shall focus henceforth on the spin zero representation  which is infinite dimensional with no highest/lowest weight states, and thus it is highly non-trivial.
In this case, the radial solution $\vv_0({\bf x}, \pi,\rho,s)$ is of the form\footnote{In the present work the solutions are shifted related to the ones found in \cite{DI3}. More precisely, let $u_k,\ w_k$ be the notation used here, and $u^{(pr)}_k,\ w^{(pr)}_k$ the notation used in \cite{DI3}. Then
$$
u^{(pr)}_k \equiv u_{k+1}, ~~~~~~~w_k^{(pr)} \equiv w_{k+1}.
$$}
\be
\vv_0({\bf x}, \pi,\rho,s)=\sum_{k=-\infty}^{-1}\,\rho^{k-1}\left(w_k\,\sqrt{\pi}+\fr{u_{k+1}}{\sqrt{\pi}}\right),
\ee
where $w_k(r,s),\ u_k(r,s)$ are expressed in terms of the first kind Kummer functions (see \cite{DI3} for more details).

The solution (\ref{gen}) of the generic non-compact Weyl equation is thus given by
\be
\vv\left({\bf x},\eta,\xi,s\right)=\fr{1}{\sqrt{\eta}}\sum_{k=-\infty}^{-1}\left(\fr{\xi+\fr{z+r}{w}}{\xi+\fr{z-r}{w}}\right)^{k-1}
\left[w_k\left(\eta+\fr{z+r}{w}\right)+u_{k+1}\left(\eta+\fr{z-r}{w}\right)\right]. \label{solgen}
\ee

Next we need to construct an infinite dimensional orthonormal basis of the aforementioned solutions.
Consider now an infinite basis of solutions $\vv_j$ given by (\ref{solgen}), where the functions $w_k$ and $u_k$ are replaced by  $w_k^{(j)}$ and  $u_k^{(j)}$, which contain certain constants of integration that can be determined via the orthonormality condition (see also \cite{DI2}):
\be
\int_{-\infty}^{1} ds\ <\vv_i, \vv_j> = \delta_{ij}.
\ee
we also refer the reader to equations (2.15), (2.17) in \cite{DI3}.

We thus need to determine the inner product $<\vv_i, \vv_j>$. Using the spherical coordinates and setting $\xi=e^{i\al}$ and $\eta=e^{i\bt}$ for $\alpha,\ \bt \in [0, 2\pi]$; the inner product of these solutions is given by
\bea
\!\!\!\!\!\!\!\!\!\!\!\!\!\!<\vv_i,\vv_j>\!\!&\!\!=\!\!&\!\fr{1}{\pi}\sum_{k,l=-\infty}^{-1}\int_0^{2\pi}
d\al\left(\fr{1+\kappa_1e^{-i(\al+\phi)}}{1+\kappa_2e^{-i(\al+\phi)}}\right)^{k-1}
\left(\fr{1+\kappa_1e^{i(\al+\phi)}}{1+\kappa_2e^{i(\al+\phi)}}\right)^{l-1} \nonumber\\
&&\hs\hs \hs \hs\hs\hs  \left(w_k^{(i)}w^{*(j)}_l \fr{1}{1-\cos\theta}+u_{k+1}^{(i)}u_{l+1}^{*(j)}
\fr{1}{1+\cos\theta}\right)\nonumber\acc
\!\!&\!\!=\!\!&\!\fr{2}{\sin^2\theta}\sum_{k,l=-\infty}^{-1}\left(\fr{\kappa_1}{\kappa_2}\right)^{k-1}\left[w_k^{(i)}w^{*(j)}_l \left(1+\cos\theta\right)+u_{k+1}^{(i)}u_{l+1}^{*(j)}\left(1-\cos\theta\right)\right], \label{ntrivial}
\eea
where $\kappa_1=\fr{1+\cos\theta}{\sin\theta}$,  $\kappa_2=\fr{1-\cos \theta}{\sin \theta}$ and $\phi \in[0,2\pi]$.

The derivation of the solutions of the generic non-compact Weyl equation and their inner product is the first step towards the construction of infinite  monopole configurations in accordance to (\ref{fields}).
Then one has to verify that these solutions satisfy the Bogomolny equations and thus, show that they correspond to BPS monopole configurations. The construction of the related fields as well as the compatibility of the fields with the Bogomolny equations are quite intricate tasks, and will be left for future investigations.

\section{Discussion}

The main results of the present investigation are: the derivation of solutions of the full generic Weyl equation in the spherically symmetric case, and the formulation of formal expressions of the associated inner products of the emerging solutions.
The extraction of the generic solutions involves suitable transformations along  the lines described in \cite{DI2} as well as a change of  variables. These results provide an essential first step towards the construction of infinite BPS monopole configurations, and as such are of great consequence. Some interesting comments are in order here.

$\bullet$ \
Let us first point out that in the $SU(2)$ ($S=0$) conventional  Weyl equation \cite{DI2} the  Nahm data are trivial and equal to $\tau_i=0$.
In the Appendix, we present the finite  $SU(n+1)$ for $n=2S+1$ case, and show that by {\it naively} setting $S=0$ in (\ref{ii}) we end up to the simple expression (\ref{trivial}) for the inner product.
However, this is not the case for the spin zero representation of the non-compact case, where the inner product as shown in (\ref{ntrivial}) involves sums of infinite number of terms and the formalism is totally different.
In particular, from the analysis of section 2 and and our recent work \cite{DI3}, it is clear that we exploit the highly non-trivial nature of the  spin zero representation as an infinite dimensional representation.
Note that, this representation lacks what are known as highest/lowest weight states, and thus it is infinite dimensional.

$\bullet$\ The $S \to \infty$ case also merits careful investigation. Although this is an interesting study we have mainly focused here on the spin zero infinite dimensional representation expressed via certain differential operators. In any case the formal similarities between the two infinite dimensional representation are indeed quite striking.

It was observed in earlier works (see e.g. \cite{floratos,W1,GCP}) that suitable rescaling of the algebra generators as $S \to \infty$ leads to a replacement of the quantum commutator with a Poisson bracket. More precisely, by setting
\be
\tau_i \to n\ \tau_i
\ee
$n =2S+1 \sim {1\over \hbar} \to \infty $, we obtain:
\be
{1 \over \hbar} \Big \{\ ,\ \Big \} = \Big [\ ,\ \Big ]
\ee
and thus the Nahm data, and the Weyl equation are accordingly modified.

 In our investigation although we deal with an infinite dimensional algebra we do not resort to Poisson algebraic structure, but we keep the quantum commutator intact within the Nahm equations.

The solutions arising in the case of the spin zero representation will be exploited to provide the Higgs and gauge fields associated to an infinite BPS monopole configuration that should satisfy the Bogomolny equation. This construction together with the validity check of these configurations with respect to the Bogomolny equation are intricate issues and will be investigated in detail in a forthcoming publication.

\appendix

\section{The $SU(n+1)$ Finite Case}
On the finite representation of dimension $n=2S+1$  the solution $\vv_0({\bf x}, \pi,\rho,s)$ has been obtained in \cite{DI3} and is of the form
\be
\vv_0({\bf x}, \pi,\rho,s)=\sum_{k=1}^n\,\rho^{k-1-S}\left(w_k\,\sqrt{\pi}+\fr{u_{k+1}}{\sqrt{\pi}}\right)
\label{fg0}
\ee
where $w_k=w_k(r,s)$ and $u_k=u_k(r,s)$ are functions of the radial coordinate and the variable $s$ only, and are given in terms of the first kind Kummer functions.

Then the orthogonal basis of the solutions (\ref{gen}) consists of the functions:
\be
\!\!\!\vv_i=\!\sum_{k=1}^n\fr{1}{\xi^S\sqrt{\eta}}\left(\xi+\fr{z+r}{w}\right)^{k-1}\!\!\left(\xi+\fr{z-r}{w}\right)^{2S-k+1}\!
\left[w_k^{(i)}\left(\eta+\fr{z+r}{w}\right)+u_{k+1}^{(i)}\left(\eta+\fr{z-r}{w}\right)\right].\label{ort}
\ee
Using the spherical coordinates and integrating on a unit circle the inner product (\ref{inner}) of the solutions (\ref{ort}) simplifies to:
\bea
\!\!<\vv_i,\vv_j>\!\!&\!\!=\!\!&\!\!\fr{2^{2S}}{\pi}\!\!\int\!\! d\al\!\sum_{k, l=1}^{n}\left(\fr{1-\cos(\al+\phi)\sin\theta}{1+\cos\theta}\right)^{2S}
\left(\fr{1+\kappa_1e^{-i(\al+\phi)}}{1+\kappa_2e^{-i(\al+\phi)}}\right)^{k-1}
\left(\fr{1+\kappa_1e^{i(\al+\phi)}}{1+\kappa_2e^{i(\al+\phi)}}\right)^{l-1}  \nonumber\acc
&&\hs \hs\hs\hs\  \left[w_k^{(i)}w_l^{*(j)}\left(\fr{1}{1-\cos\theta}\right)+
u_{k+1}^{(i)}u_{l+1}^{*(j)}\left(\fr{1}{1+\cos\theta}\right)\right].\nonumber\\ \label{ii}
\eea

These results should recover the ones obtain in \cite{DI2}.
Indeed, we can confirm that by concentrating on some specific examples.
In particular, in the $SU(2)$ and $SU(3)$ cases we have:

$\bullet$ {\bf $  SU(2)$ Case}. Here $n=1$ therefore, $S=0$;   while the inner product (\ref{ii}) is equal to
\be
<\vv_i,\vv_j>=\fr{2}{\sin^2\theta}\left[w_1^{(i)}w_1^{*(j)}\left(1+\cos\theta\right)+u_1^{(i)}u_1^{*(j)}
\left(1-\cos\theta\right)\right]. \label{trivial}\acc
\ee

$\bullet$ {\bf $SU(3)$ Case}. Here $n=2$  and thus, $S=\fr{1}{2}$; while the inner product (\ref{ii}) is equal to
\be
\!\!\!<\vv_i,\vv_j>=\fr{4}{\sin^2\theta}\!\left[w_1^{(i)}w_1^{*(j)} +u_1^{(i)}u_1^{*(j)} \left(\fr{1-\cos\theta}{1+\cos\theta}\right)
+w_2^{(i)}w_2^{*(j)}\left(\fr{1+\cos\theta}{1-\cos\theta}\right)+u_2^{(i)}u_2^{*(j)} \right].
\ee
The specific forms of the functions $w_k^{(i)}$ and $u_k^{(i)}$ have been obtained in \cite{DI,DI2} and are given in terms of the Whittaker $M$ functions\footnote{The first kind Kummer functions $M_k$ obtained in \cite{DI3} are equivalent to the Whittaker $M$ functions obtained in \cite{DI} due to the relation
$$
M(k, \mu ; z) = e^{-{z\over 2}}\ z^{\mu +{1\over 2}}\ M_{k}(\mu -k+{1\over 2}, 1+2 \mu;z).
$$}.

\vskip 20pt
\centerline{\bf Acknowledgements}
\vspace{.25cm}
\noindent

We are indebted to J. Avan for illuminating discussions.

\end{document}